\documentclass{article}
\usepackage{spconf,amsmath,graphicx,amssymb}
\usepackage{mathabx}
\usepackage[utf8]{inputenc}
\usepackage[english]{babel}
\usepackage{amsmath,amsfonts,amssymb,amsthm}
\newtheorem{thm}{Theorem}
\newtheorem{example}[thm]{Example}
\usepackage{mathtools}
\usepackage[footnotesize]{caption}
\usepackage{subcaption}
\PassOptionsToPackage{hyphens}{url}\usepackage{hyperref}

\title{CONTRASTIVE SELF-SUPERVISED LEARNING FOR WIRELESS POWER CONTROL}
\name{Navid Naderializadeh}
\address{HRL Laboratories, LLC\\ Malibu, California, USA}
\begin{document}
%
\maketitle
\begin{abstract}
We propose a new approach for power control in wireless networks using self-supervised learning. We partition a multi-layer perceptron that takes as input the channel matrix and outputs the power control decisions into a backbone and a head, and we show how we can use contrastive learning to pre-train the backbone so that it produces similar embeddings at its output for similar channel matrices and vice versa, where similarity is defined in an information-theoretic sense by identifying the interference links that can be optimally treated as noise. The backbone and the head are then fine-tuned using a limited number of labeled samples. Simulation results show the effectiveness of the proposed approach, demonstrating significant gains over pure supervised learning methods in both sum-throughput and sample efficiency.\footnote{Code available at \url{https://github.com/navid-naderi/ContrastiveSSL_WirelessPowerControl}.}
\end{abstract}
\begin{keywords}
Interference management, power control, self-supervised learning, contrastive learning
\end{keywords}
\section{Introduction}
\label{sec:intro}

Machine learning (ML), and especially deep learning, methods have shown very promising performance in many domains, where the main challenge is to implicitly or explicitly model the underlying input distribution or the input/output relationship in the data. Among many areas that have benefited from the breakthroughs in ML, wireless communications is one of the most prominent ones, where ML has been used to tackle various problems, including network optimization and radio resource management (RRM)~\cite{chen2019artificial}.


Under the ML-based RRM paradigm, earlier works focused on supervised approaches~\cite{sun2017learning}. In these methods, ground-truth labels for RRM decisions were created using baseline algorithms, e.g., weighted minimum mean-squared error (WMMSE)~\cite{shi2011iteratively}, that had been shown to work well across a variety of scenarios, but exhibited lower complexity than solving the original problems. Nevertheless, deriving optimal solutions for RRM problems is challenging as the corresponding optimization problems are typically very complex~\cite{besser2020deep}. This has given rise to a plethora of work using other types of ML for RRM, such as reinforcement learning and unsupervised learning, which do not necessarily rely on manual labeling of the data for training the underlying neural networks~\cite{nasir2019multi, shen2019graph, spawc2020_DRL_RRM, eisen2019learning, eisen2020optimal, 9154336}.

Another paradigm in ML that has recently gained significant traction is self-supervised learning (SSL). The core idea of SSL is that the underlying neural network parameters can be pre-trained using labels that are generated manually from the data. For example, in case of visual data, an input image can be manually rotated, and the network can then be trained to predict the rotation angle~\cite{gidaris2018unsupervised}. SSL has recently shown striking success in representation learning in computer vision and natural language processing~\cite{kolesnikov2019revisiting, Lan2020ALBERT:}.


In this paper, we use the notion of contrastive learning~\cite{oord2018representation, chen2020simple, wang2020understanding} for injecting SSL into learning optimal RRM decisions. We first break the underlying multi-layer perceptron (MLP) into a \emph{backbone}, which maps the input channel matrix to an intermediate \emph{embedding}, and a regression \emph{head}, which maps the embedding to an output power control vector. We first pre-train the backbone using contrastive loss, which forces the embeddings of two \emph{similar} channel matrices to be close, while the embeddings of two \emph{dissimilar} channel matrices are pushed farther from each other. We use the information-theoretic condition for the optimality of treating interference as noise~\cite{geng2015optimality, naderializadeh2014itlinq} to identify and randomly remove weak interference links across the network to create similar channel matrices. Once pre-training is complete, the regression head can then be trained alongside the backbone using only a \emph{limited} set of labeled training samples. To the best of our knowledge, this is the first work that uses contrastive SSL in the wireless communications domain.

We show that the proposed self-supervised pre-training, coupled with the aforementioned information-theoretic channel augmentations, can effectively partition the embedding space into clusters within each of which the power control decisions are similar. Note that such a clustering happens without any supervision, i.e., the use of ground-truth power control decisions. We further demonstrate how the subsequent supervised training phase can train the regression head to produce power control decisions with a much fewer number of labeled samples compared to a pure supervised baseline.

\section{Problem Formulation}\label{sec:sys_model}

We consider a wireless network with $N$ transmitter-receiver pairs $\{(\mathsf{Tx}_i, \mathsf{Rx}_i)\}_{i=1}^N$, where each transmitter $\mathsf{Tx}_i$ intends to send a message to its corresponding receiver $\mathsf{Rx}_i$. The transmissions are assumed to occur using the same time and frequency resources, implying that simultaneous transmissions of two or more transmitters would cause interference on each other. Let $x_i$ denote the signal transmitted by transmitter $\mathsf{Tx}_i$, subject to a transmit power constraint of $P_{\max}$. Then, the received signal of receiver $\mathsf{Rx}_i$ can be written as
\begin{align}
y_i = h_{ii} x_i + \sum_{j\neq i} h_{ij} x_j + n_i,
\end{align}
where $h_{ij}$ denotes the channel gain between transmitter $\mathsf{Tx}_j$ and receiver $\mathsf{Rx}_i$, and $n_i\sim\mathcal{CN}(0,\sigma^2)$ denotes the additive white Gaussian noise at receiver $\mathsf{Rx}_i$. Letting $P_i\in[0, P_{\max}]$ denote the transmit power used by transmitter $\mathsf{Tx}_i$, the Shannon capacity of the channel between transmitter $\mathsf{Tx}_i$ and receiver $\mathsf{Rx}_i$ can be written as
\begin{align}
R_i &= \log_2\left(1 + \tfrac{|h_{ii}|^2 \gamma_i}{\sum_{j\neq i}|h_{ij}|^2 \gamma_j + \frac{\sigma^2}{P_{\max}}}\right),
\end{align}
where $\gamma_i=\frac{P_i}{P_{\max}}\in[0,1]$ denotes the fraction of the maximum transmit power used by transmitter $\mathsf{Tx}_i$.

In this paper, we focus on the problem of power control for sum-rate maximization across the network. In particular, we intend to solve the following optimization problem:
\vspace{-.1in}
\begin{subequations}\label{eq:opt_PC}
\begin{alignat}{2}
&\max_{\gamma_1,...,\gamma_N}&& \quad \sum_{i=1}^N R_i \\
&\quad~\text{s.t.} && \quad R_i = \log_2\left(1 + \tfrac{|h_{ii}|^2 \gamma_i}{\sum_{j\neq i}|h_{ij}|^2 \gamma_j + \frac{\sigma^2}{P_{\max}}}\right), ~\forall i \\
& && \quad \gamma_i\in[0, 1], ~\forall i.
\end{alignat}
\end{subequations}


\section{Proposed Approach}

We propose solving problem~\eqref{eq:opt_PC} using a learning-based approach. In particular, we consider a multi-layer perceptron (MLP), represented as a mapping $\mathsf{MLP}: \mathbb{R}^{N\times N} \rightarrow [0,1]^{N}$, which takes as input the $N\times N$ channel matrix
\begin{align}
\mathbf{H} \coloneqq
\begin{bmatrix}
|h_{11}|^2 & \dots & |h_{1N}|^2 \\
\vdots & \ddots & \vdots \\
|h_{N1}|^2 & \dots & |h_{NN}|^2 \\
\end{bmatrix},
\end{align}
flattened into an $N^2$-dimensional vector, and generates the power control vector $\mathbf{\Gamma} = \begin{bmatrix} \gamma_1 & \dots & \gamma_N\end{bmatrix}^T$ at the output.

In order to train this MLP, we need a training dataset comprising multiple samples, each with an input channel matrix $\mathbf{H}$ and the corresponding output power control vector $\mathbf{\Gamma}$. However, the optimization problem~\eqref{eq:opt_PC} is non-convex and challenging to solve exactly, especially as the number of transmitter-receiver pairs grows. To circumvent that issue, similar to~\cite{sun2017learning}, we resort to deriving sub-optimal power control decisions using WMMSE~\cite{shi2011iteratively}. Nevertheless, running the WMMSE algorithm itself to create a massive dataset is computationally expensive, which justifies using both \emph{labeled} samples, i.e., $(\mathbf{H},\mathbf{\Gamma})$ pairs, as well as \emph{unlabeled} samples, only including an input channel matrix $\mathbf{H}$.

More formally, we assume that we have access to a set of $M$ training samples, which consist of $M_L$ labeled samples $\{(\mathbf{H}_i,\mathbf{\Gamma}_i)\}_{i=1}^{M_L}$, alongside $M_U = M-M_L$ unlabeled samples $\{\mathbf{H}_j\}_{j={M_L+1}}^{M}$. Note that generating the input channel matrices is much less computationally expensive than deriving the WMMSE power control decisions, so it can be the case that $M_U \gg M_L$, which represents the so-called ``low-label'' regime.

As in most prior work in the area of wireless power control based on supervised learning, e.g.,~\cite{sun2017learning}, the $M_L$ labeled samples can be used to train the MLP in an end-to-end manner using stochastic gradient descent (SGD). In particular, using a batch of samples $\{(\mathbf{H}_i,\mathbf{\Gamma}_i)\}_{i\in\mathcal{B}_L}, \mathcal{B}_L\subseteq\{1,\dots,M_L\}$, the regression objective minimized via SGD can be written as
\begin{align}\label{eq:supervised_loss}
\mathcal{L}_{\mathsf{supervised},\mathcal{B}_L} = \frac{1}{|\mathcal{B}_L|} \sum_{i\in\mathcal{B}_L} \big\| \mathsf{MLP}(\mathbf{H}_i) - \mathbf{\Gamma}_i \big\|_2^2
\end{align}
which denotes the mean-squared error (MSE) between the MLP outputs and the ground-truth power control decisions.

In the following, we will demonstrate how we can use all the labeled and unlabeled training samples to \emph{pre-train} the MLP parameters using self-supervised learning.

\begin{figure*}[t]
\includegraphics[width=\textwidth]{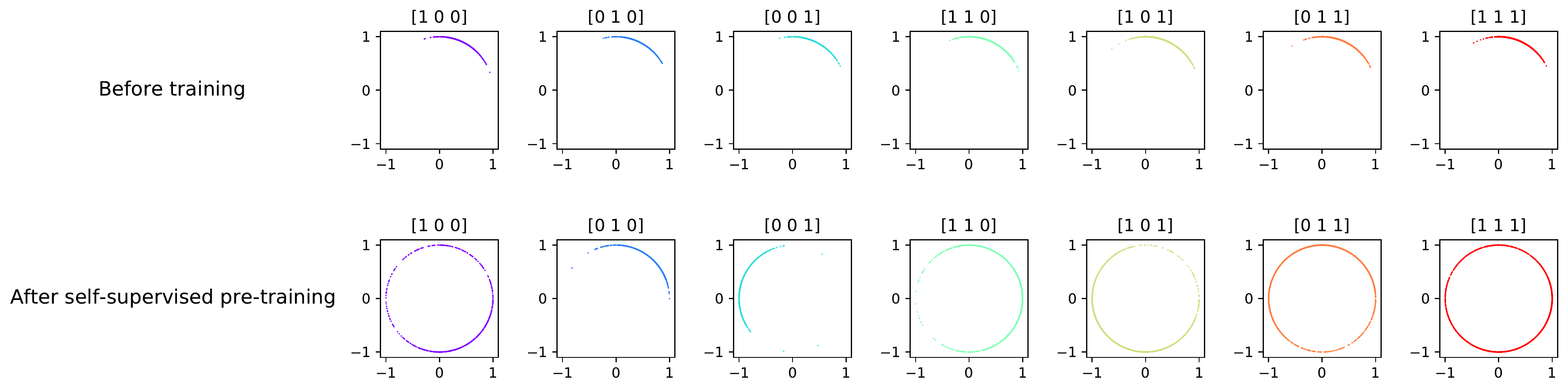}
\caption{Embeddings of channel matrices before/after contrastive self-supervised pre-training for a network with $N=3$ transmitter-receiver pairs. Each column illustrates the embeddings of channel matrices with similar optimal power control decisions. Each binary vector represents the on/off patterns of the transmitters, which is closest to the output of the WMMSE algorithm~\cite{shi2011iteratively}. Note that these clusters emerge solely based on the channel matrices and the corresponding augmentations satisfying~\eqref{eq:cond_sufficient_TIN}, without the use of any ground-truth power control decisions.}
\label{fig:ssl_clusters}
\end{figure*}

\subsection{Self-Supervised Pre-Training}\label{sec:SSL}
Let $l$ denote the number of neurons in the last hidden layer of the MLP, right before the $N$-dimensional power control decision outputs. The MLP can, therefore, be decomposed into a \emph{backbone} $f: \mathbb{R}^{N\times N} \rightarrow \mathbb{R}^{l}$ and a \emph{head} $g: \mathbb{R}^{l} \rightarrow [0,1]^{N}$, where for any input channel matrix $\mathbf{H}\in\mathbb{R}^{N\times N}$,
\begin{align}
\mathsf{MLP}(\mathbf{H}) = g \circ f(\mathbf{H}).
\end{align}
We call $f(\mathbf{H})$ the \emph{embedding} of the channel matrix $\mathbf{H}$. This embedding is an intermediate representation of the channel matrix, which is implicitly learned as part of the supervised learning process. However, here we show that this embedding can be pre-trained without the need for any labels, i.e., ground-truth WMMSE power control decisions.

We use the idea of contrastive learning~\cite{oord2018representation, chen2020simple, wang2020understanding}, which states that the embeddings of two \emph{similar} channel matrices should end up close to each other in the $l$-dimensional embedding space, while the embeddings for two \emph{different} channel matrices should get as far as possible from each other. In particular, consider a set of channel matrix tuples $\{(\overline{\mathbf{H}}_i, \underline{\mathbf{H}}_i)\}_{i\in\mathcal{B}}$, where for each $i\in\mathcal{B}$, $\overline{\mathbf{H}}_i$ and $\underline{\mathbf{H}}_i$ are similar, while for any $j\neq i$, $\overline{\mathbf{H}}_i$ is different from $\underline{\mathbf{H}}_j$. Then, the backbone $f(\cdot)$ can be trained so as to minimize the following \emph{contrastive} loss,
\vspace{-.1in}
\begin{align}\label{eq:contrastive_loss}
&\mathcal{L}_{\mathsf{contrastive},\mathcal{B}} = \nonumber\\ &\quad\frac{1}{|\mathcal{B}|}\sum_{i\in\mathcal{B}}\left[ -\log \frac{e^{\frac{f(\overline{\mathbf{H}}_i)^T f(\underline{\mathbf{H}}_i)}{\tau}}}{e^{\frac{f(\overline{\mathbf{H}}_i)^T f(\underline{\mathbf{H}}_i)}{\tau}} + \sum_{j\neq i} e^{\frac{f(\overline{\mathbf{H}}_i)^T f(\underline{\mathbf{H}}_j)}{\tau}}} \right],
\end{align}
where $\tau$ denotes a scalar \emph{temperature} hyperparameter. In order to implement the contrastive loss in~\eqref{eq:contrastive_loss}, we use mini-batch gradient descent on \emph{both} labeled and unlabeled samples, i.e., $\{\mathbf{H}_i\}_{i=1}^{M}$. In particular, given a mini-batch $\{\mathbf{H}_i\}_{i\in\mathcal{B}}$, where $\mathcal{B}\subseteq\{1,\dots,M\}$, for each channel matrix $\mathbf{H}_i$, we create two \emph{augmentations} $\overline{\mathbf{H}}_i$ and $\underline{\mathbf{H}}_i$, treating them as similar samples, while we use an augmentation of the rest of the channel matrices in the mini-batch for the dissimilar samples.

To create the channel matrix augmentations, we leverage the information-theoretic optimality condition for treating interference as noise~\cite{geng2015optimality}. Theorem 4 in~\cite{geng2015optimality} states that in a Gaussian interference channel as modeled in Section~\ref{sec:sys_model}, treating interference as noise is information-theoretically optimal (to within a constant gap of the capacity region) if
\begin{align}\label{eq:tin_opt_orig}
\frac{P_{\max} |h_{ii}|^2}{\sigma^2} \geq \frac{P_{\max} |h_{ij}|^2}{\sigma^2} \cdot \frac{P_{\max} |h_{ki}|^2}{\sigma^2}, \forall i, \forall j,k\neq i.
\end{align}
As noted in~\cite{naderializadeh2014itlinq}, it is easy to verify that a sufficient condition for the condition in~\eqref{eq:tin_opt_orig} would be that
\begin{align}\label{eq:cond_sufficient_TIN}
\frac{P_{\max} |h_{ij}|^2}{\sigma^2} \leq 
\sqrt{\frac{P_{\max} \min(|h_{ii}|^2,|h_{jj}|^2)}{\sigma^2}}, \forall j\neq i.
\end{align}
Indeed, if an interference link between transmitter $\mathsf{Tx}_j$ and a non-desired receiver $\mathsf{Tx}_i$ is so weak that it satisfies~\eqref{eq:cond_sufficient_TIN}, then that interference link might be treated as noise without the loss of optimality (again, to within a constant gap).

In light of the aforementioned conditions, given a channel matrix $\mathbf{H}$, we identify the weak interference links that satisfy~\eqref{eq:cond_sufficient_TIN}. Then, we create an augmentation of the channel matrix by keeping each weak interference link with probability 0.5 and discarding it (setting it to zero) otherwise. Therefore, two different augmentations of a given channel matrix have the identical signal links as well as \emph{strong} interference links, while their weak interfering link topologies might differ.

\begin{figure*}[t]
\includegraphics[width=\textwidth]{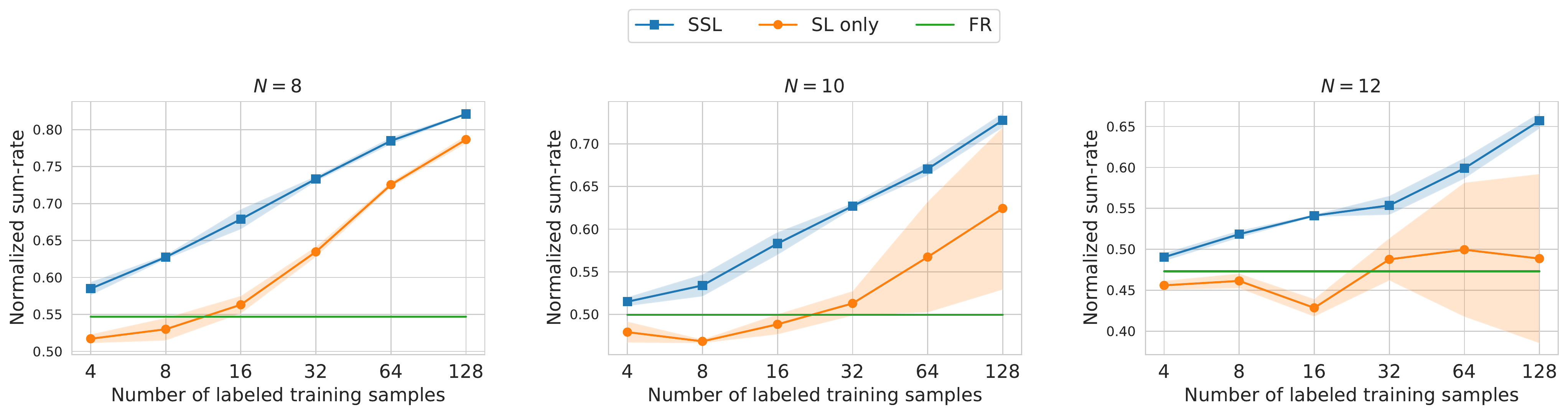}
\caption{Sum-rates achieved by our proposed SSL approach (normalized by sum-rates achieved by WMMSE) versus the number of labeled training samples, as compared to a pure supervised approach (SL only) and full reuse (FR) in networks with $N\in\{8,10,12\}$ transmitter-receiver pairs.}
\label{fig:kshot_sumrates}
\end{figure*}

To illustrate the effectiveness of such an augmentation method coupled with the contrastive self-supervised pre-training, we consider a network with $N=3$ transmitter-receiver pairs, and a backbone $f(\cdot)$ with a single $128$-dimensional hidden layer, LeakyReLU non-linearity, and an embedding size of $l=2$. As suggested in prior work on contrastive representation learning~\cite{chen2020simple, wang2020understanding, wang2017normface}, we use $\ell_2$ normalization to ensure that each input channel matrix is mapped to an embedding on the unit circle. For simplicity, we set $\frac{P_{\max}}{\sigma^2}=1$ and generate $M=10^4$ training channel matrices, where in order to capture short-term fading, each channel gain is modeled as a Rayleigh random variable with zero mean and unit variance, independent of all the other channel gains in the dataset. We use a batch size of 64, a learning rate of 0.05, and a temperature of $\tau=0.1$ in~\eqref{eq:contrastive_loss}.

Prior to the self-supervised pre-training procedure, we derive the 2-D embeddings of all the $M=10^4$ channel matrices in the training set. We then pre-train the backbone for 20 epochs using the aforementioned augmentations and the contrastive loss~\eqref{eq:contrastive_loss}, without using any of the ground-truth power control decisions. At the end of the pre-training phase, we derive the 2-D embeddings of the channel matrices again. Figure~\ref{fig:ssl_clusters} demonstrates the embeddings before/after the pre-training phase. Each column represents the embeddings of the channel matrices with similar \emph{binary} power control decisions derived using the WMMSE approach.\footnote{For each input channel matrix, we first derive the WMMSE power control decisions and we then map those power levels to the closest (in an $\ell_2$ sense) binary power control vector in $[0,1]^3$. Note that binary power control is indeed optimal in many scenarios of interest~\cite{gjendemsjo2008binary}, and the WMMSE outputs are already in binary form in a majority of the cases.} As the figure shows, prior to training, the backbone cannot distinguish dissimilar channel matrices with different power control requirements at all. However, note how well the embeddings are clustered on the 2-D unit circle embedding space after self-supervised pre-training \emph{without the use of any ground-truth power control decisions}. Quite interestingly, the clusters show a meaningful pattern. For example, the cluster of embeddings corresponding to the second transmitter being active only (2\textsuperscript{nd} column from left) complements the cluster of embeddings corresponding to the first and third transmitters being active, while the first transmitter is silent (3\textsuperscript{rd} column from right).

\subsection{Supervised Training}
Subsequent to the self-supervised pre-training phase, we can use the $M_L$ labeled samples to train the head $g(\cdot)$ together with the pre-trained backbone $f(\cdot)$ to minimize the supervised MSE loss~\eqref{eq:supervised_loss}. Note that the contrastive loss~\eqref{eq:contrastive_loss} on both the labeled and unlabeled samples can still be utilized to fine-tune the backbone in the main training phase as well. In particular, for a batch of a labeled samples $\{(\mathbf{H}_i,\mathbf{\Gamma}_i)\}_{i\in\mathcal{B}_L}, \mathcal{B}_L\subseteq\{1,\dots,M_L\}$, we minimize the aggregate loss
\begin{align}
\mathcal{L}_{\mathsf{total},\mathcal{B}_L} = \mathcal{L}_{\mathsf{supervised},\mathcal{B}_L} + \alpha_{\mathsf{contrastive}}\cdot \mathcal{L}_{\mathsf{contrastive},\mathcal{B}_L},
\end{align}
where $\alpha_{\mathsf{contrastive}}\in\mathbb{R}_+$ is a hyperparameter, while for a batch of unlabeled samples $\{(\mathbf{H}_j)\}_{j\in\mathcal{B}_U}, \mathcal{B}_U\subseteq\{M_L+1,\dots,M\}$, we simply minimize the contrastive loss
\begin{align}
\mathcal{L}_{\mathsf{total},\mathcal{B}_U} =  \mathcal{L}_{\mathsf{contrastive},\mathcal{B}_L}.
\end{align}

To evaluate the final system-level performance of the proposed approach, we consider networks with $N\in\{8,10,12\}$ transmitter-receiver pairs and $M=10^3$ training samples. We set $\alpha_{\mathsf{contrastive}}=1$, and use an MLP with two $512$-dimensional hidden layers, LeakyReLU non-linearity for the hidden layers, and sigmoid non-linearity at the output. Supervised training is run for $100$ epochs. The rest of the (hyper)parameters are the same as those in Section~\ref{sec:SSL}. For each value of $N$, once training is completed, we evaluate the performance of the trained MLP on $10^3$ test samples.

Figure~\ref{fig:kshot_sumrates} demonstrates the achieved sum-rates of our proposed approach (SSL), normalized by those achieved by WMMSE, as a function of the number of labeled samples, i.e., $M_L$. For comparison, we have included the full reuse baseline, where all transmitters use their full transmit power. Moreover, the performance of a conventional approach that only uses the labeled samples for supervised learning is also shown, for which we lower the learning rate to 0.01 to optimize the performance. The solid lines and shadowed regions respectively represent the mean and standard deviation over three training runs with different random seeds. As the figure shows, our proposed approach exhibits a significant performance boost compared to a pure supervised approach, especially in the low-label regime and for larger network sizes. In particular, for a given number of labeled samples, the proposed approach provides a higher sum-rate performance, and for a given target sum-rate, much less labeling effort is required to achieve the same performance level.

\section{Concluding Remarks}
We considered the problem of downlink power control to maximize sum-throughput in wireless networks, where a multi-layer perceptron (MLP) is trained to produce power control decisions given an input channel matrix. We proposed using a self-supervised pre-training approach, in which the backbone (the MLP excluding the final layer) is trained using contrastive loss, so that similar channel matrices are represented using similar embeddings and vice versa. We used an information-theoretic condition for the optimality of treating interference as noise to create augmentations for the input channel matrices as a basis for computing the contrastive loss. We showed how a subsequent supervised training phase can boost the resulting sum-rate of the proposed method, using many fewer labels than a pure supervised approach.

Nevertheless, two limitations of a fully-connected backbone are that i) it can only be applied to a fixed number of transmitters and receivers, and ii) it is sensitive to the ordering of the transmitter-receiver pairs, hence not permutation-equivariant. Both of these issues can be addressed by a graph neural network (GNN) backbone~\cite{shen2019graph,eisen2020optimal,9154336}. Studying the impact of GNNs on the proposed self-supervised approach is an interesting direction that we leave for future work.

\vfill\pagebreak

\bibliographystyle{IEEEbib}
\bibliography{refs}

\end{document}